\documentclass[a4paper,showpacs,superscriptaddress,nofootinbib,10pt]{revtex4}
\usepackage{amsmath,amssymb,bm,natbib}
\usepackage{epsfig}
\usepackage{graphicx}
\usepackage{multirow}
\usepackage{slashed}
\makeatletter

\newcommand{\Rmnum}[1]{\expandafter\@slowromancap\romannumeral #1@}
\makeatother
\begin{document}
\def\bib{\bibitem}
\def\be{\begin{equation}}
\def\ee{\end{equation}}
\def\beq{\begin{equation}}
\def\eeq{\end{equation}}
\def\beqar{\begin{eqnarray}}
\def\eeqar{\end{eqnarray}}
\def\barr{\begin{array}}
\def\earr{\end{array}}
\def\lsim{\:\raisebox{-0.5ex}{$\stackrel{\textstyle<}{\sim}$}\:}
\def\gsim{\:\raisebox{-0.5ex}{$\stackrel{\textstyle>}{\sim}$}\:}
\def\tilh{\tilde{h}}
\def\and{\qquad {\rm and } \qquad}
\def\vev{{\it v.e.v. }}
\def\p{\partial}
\def\ga{\gamma^\mu}
\def\slp{p \hspace{-1ex}/}
\def\sleps{ \epsilon \hspace{-1ex}/}
\def\slk{k \hspace{-1ex}/}
\def\slq{q \hspace{-1ex}/\:}
\def\prl#1{Phys. Rev. Lett. {\bf #1}}
\def\prd#1{Phys. Rev. {\bf D#1}}
\def\plb#1{Phys. Lett. {\bf B#1}}
\def\npb#1{Nucl. Phys. {\bf B#1}}
\def\mpl#1{Mod. Phys. Lett. {\bf A#1}}
\def\ijmp#1{Int. J. Mod. Phys. {\bf A#1}}
\def\zp#1{Z. Phys. {\bf C#1}}
\def\etal{ {\it et al.} }
\def\ie{ {\it i.e.} }
\def\eg{ {\it e.g.} }
\def\sbar{ \overline{s} }
\def\thmin{\theta_0}
\def\cmin{\cos \theta_0}
\def\pepebar{P_{e} P_{\overline{e}}}
\def\eebar{$e^+e^-~$}
\def\eegz{$e^+e^- \to \gamma Z$}
\def\ggz{$\gamma\gamma Z~$}
\def\gzz{$\gamma ZZ~$}

\title{
Isolating CP-violating $\gamma ZZ$ coupling 
in $e^+e^- \to \gamma Z$ 
with transverse beam polarizations
}
\author{{\bf B. Ananthanarayan}}
\author{{\bf Sumit K. Garg}}
\author{{\bf Monalisa Patra}}
\affiliation{
Centre for High Energy Physics, 
Indian Institute of Science, 
Bangalore 560 012, India} 

\author{{\bf Saurabh D. Rindani}}

\affiliation{ 
Theoretical Physics Division,
Physical Research Laboratory,
Navrangpura, Ahmedabad 380 009, India}


\begin{abstract}
We revisit the process $e^+e^- \to \gamma Z$ at the ILC with transverse
beam polarization in the presence of anomalous CP-violating 
$\gamma Z Z$ coupling $\lambda_1$ and $\gamma \gamma Z$ coupling
$\lambda_2$.  We point 
out that if the final-state spins are resolved, then it becomes possible to 
fingerprint the anomalous coupling {\rm Re}$\lambda_1$. 
90\% confidence level limit
on {\rm Re}$\lambda_1$ achievable at ILC with center-of-mass energy of
500 GeV or 800 GeV with realistic initial beam
polarization and integrated luminosity is 
of the order of few times of $10^{-2}$ 
when the helicity of  $Z$ is used and
$10^{-3}$ when the helicity of $\gamma$ is used.  The resulting 
corrections at quadratic order
to the cross section and its influence on these limits are also
evaluated and are shown to be small.
The benefits of such  polarization programmes at the ILC are compared 
and contrasted for the process at hand.
We also discuss possible methods by
which one can isolate events with a definite helicity for one of the 
final-state
particles. 
\end{abstract}

\pacs{13.66.-a, 12.60.-i, 13.88.+e}
\maketitle

\section{Introduction}

The International Linear Collider (ILC)~\cite{ILC1,ILC2} is a proposed
collider that will collide electrons and positrons at high energy and
luminosity and is expected to verify the predictions of the
standard model (SM) at a high level of precision, 
and to establish interactions beyond the SM even 
if there is no direct
production of particles that are not in the spectrum of
the SM.  One window to such new physics is the
discovery of CP violation beyond what is predicted
by the SM; for a review on basic principles of
CP violation at colliders, see Ref.~\cite{SDR}.  
It has been shown that
the availability of beam polarization of one or both of the beams,
transverse and longitudinal,
can significantly enhance the sensitivity to such beyond the
SM interactions; see, e.g., Ref.~\cite{review} for a recent review.

If  beyond the SM interactions are very subtle, then it is
particularly important to search for deviations from the SM
to linear order.
Keeping these considerations in mind, 
the issue of sensitivity to linear order of CP-violating anomalous gauge
boson couplings, denoted by $\lambda_1$ ($\gamma Z Z $) and $\lambda_2$
($\gamma \gamma Z$) to be
defined later, that
contribute to the process $e^+e^-\to \gamma Z$
with longitudinal beam polarization was considered in Ref.~\cite{CR},
and with transverse beam polarization in
Ref.~\cite{ARSB} (for earlier discussions in the context of
unpolarized beams
see Ref.~\cite{hagiwaraetal,bauretal,gounarisetal}, and Ref.~\cite{Renard}
for beam polarization effects).  These couplings 
are absent in the SM even at loop level, and 
can be thought of  representing some basic 
interactions arising from an underlying theory;
for instance they can arise in some
extensions of the SM at the one-loop level \cite{ext1,ext2,ext3,ext4}, 
and thus their nonvanishing 
value indicates a signature of new physics. 
Anomalous couplings have been investigated at the Large Electron
Positron (LEP) collider, resulting in limits of the order of 0.05
(0.13) on the magnitude of $\gamma \gamma Z$ ($\gamma ZZ$) couplings
\cite{lepewwg}.
The most stringent bounds on the absolute value of these couplings 
comes from the 
recent D0 and CDF Collaboration results
of Tevatron~\cite{explimits1, explimits2}.
Apart from these direct
limits, imposing unitarity of partial wave scattering amplitudes can
give limits on the couplings \cite{bauretal,czyz}. 
For an ILC operating at $\sqrt{s}=500$ GeV, the analysis of Czyz et al.
\cite{czyz} gives the unitarity limits
$\vert \lambda_{1,2} \vert \lsim 2$.
Other unitarity limits are expressed as 
of the order of (0.1 TeV$^3)/\Lambda^3$ on dimension-6 couplings and
of the order of ($2\cdot 10^{-3}$ TeV$^5)/\Lambda^5$ on dimension-8
couplings, where $\Lambda$ is the assumed scale of new physics
\cite{bauretal}.  A review of these results may be found in
Ref.~\cite{RPP}.  Considerations of the anomalous sector involving
W bosons have also been recently studied in the
context of the Large Hadron Collider, see Refs.~\cite{WW1,WW2}
and Ref.~\cite{WW3} for quartic couplings involving
$\gamma$ pairs along with W pairs as well as Z pairs
and references therein.

Since, in the  process at hand, $t$- and $u$-channel exchanges are present
there is an additional dependence on the polar scattering angle ($\theta$)
between $\gamma$ and $e^-$ direction.
A polar-angle forward-backward asymmetry is seen 
with longitudinally-polarized beams due to the interference of the CP-violating anomalous coupling 
with SM contribution~\cite{CR}, because the photon should be 
produced symmetrically if CP is conserved. Longitudinal beam polarization
improves sensitivity to some of the form factors,  whereas the transversely
polarized beams with new combinations of polar and azimuthal asymmetries
enable better measurement. This is due to the fact that with transverse 
beam polarization, one 
has an additional angle $\phi$, the azimuthal angle
with respect to the direction of quantization of the electron
polarization, which allows one to obtain distributions that
are sensitive to the real parts of the anomalous couplings.
In order to obtain the fully differential cross sections 
in the presence of anomalous gauge couplings, one may use 
the helicity amplitudes listed in Ref.~\cite{czyz} and
account for the transverse polarization using the
generalized formalism of Ref.~\cite{hikasa}.  
Suitable asymmetries may then be constructed which can be
used to extract these anomalous couplings.  Furthermore,
one can obtain  90\% confidence limits on these 
couplings if no signal is observed for realistic beam
polarization and typical integrated luminosities.
It may be emphasized that the limits obtained in this
work are completely based on an analytical approach and
is a strength of the method and is an extension of
the approach of Ref.~\cite{ARSB} which have
proven necessary for the extraction of
Re~$\lambda_1$.  It would be a useful benchmark
for future simulation studies of the same system.

As a consequence
of the CPT theorem, in the absence of beam polarization and
with longitudinal beam polarization, only the imaginary
parts of these contribute to the cross section at linear order.
It turns out that with transverse beam polarization and
due to the interference of the SM amplitudes with those
arising from anomalous couplings, only
{\rm Re}$\lambda_2$ contributes to the
fully differential cross section, whereas the {\rm Re}$\lambda_1$
contribution vanishes because the photon has only vectorlike
couplings.  Thus the question of isolating ${\rm Re}\lambda_1$
to leading order remains open.

Recently, it was shown, in the context of $t\bar{t}$ production
with beyond the SM interactions parametrized in terms of
effective four-Fermi interactions, that the measurement of 
final-state helicity can help in disentangling the contributions of
scalar and tensor like four-Fermi interactions when the beams
are transversely polarized~\cite{APR}.

Inspired by the considerations in the work above, we now ask
whether at leading order one can isolate ${\rm Re}\lambda_1$ by
the measurement of the helicity of the $Z$ or the photon.
The answer is in the affirmative.  Whereas the considerations
of the top-quark helicity referred to above can be extended to
the $Z$, which we describe in a little detail later on, there
is no analogous method for determining the helicity of the
photon as it is a stable particle.  At low energies the final-state 
helicity determination is a
key ingredient for the determination of neutrino helicity in
the well-known Goldhaber experiment~\cite{Goldhaber}. However, it 
is conceivable that there are materials that can be used in the construction
of the ILC detectors which could be used for these determinations.
In particular aligned crystals could be candidates for such detectors 
in case of 
high-energy photons, as has been considered by the NA59 
collaboration\cite{Na59Collb}.
Leaving aside the experimental question of measurement of
the final-state helicities, which would affect the sensitivity,
we construct  asymmetries that
involve, e.g., samples of positive and negative $Z$ helicities,
and also photon helicities.  These are then translated into
90\% confidence limits on the anomalous couplings ${\rm Re}\lambda_{1,2}$
by combining them with the previously established results.
The best limits are expressed in the ${\rm Re}\lambda_1$-${\rm
Re}\lambda_2 $ plane.  If these are set to zero one at a time,
then we obtain the limit on ${\rm Re}\lambda_1$ to be 0.0958
when the helicity of only the $Z$ is resolved for a total center-of-mass
energy (c.m.) of 500 GeV with an integrated luminosity of 500 fb$^{-1}$ and
realistic polarizations.  An improved limit of $0.0034$
is obtained  when the helicity of $\gamma$ is resolved. 
The stability of these limits
when they are fed back into the expressions at quadratic order for
the cross section is also addressed, by iteratively including the
effects  which turn out to be not of great significance in practice.

This paper is organized as follows:  In Sec.~\ref{formalism},
we recall the basic vertices and present the definitions,
followed by Sec.~\ref{distributions}, where we present the
results of our computation of the fully differential cross section
with final-state helicity resolution.  In Sec.~\ref{asymmetries}
we define the asymmetries that we have used to obtain 90\% confidence
limits.  In Sec.~\ref{discussion} we present a detailed discussion
on the possibility of obtaining samples of define helicities and
discuss the issue of photon and $Z$ helicity measurement.  In Sec.~\ref{summary}
we present a summary of the results and a discussion.

\section{Formalism for the process \eegz}\label{formalism}
We begin by writing down the formalism for the process closely
following the treatment in Ref.~\cite{ARSB}.
We consider the process 
\begin{equation}
e^-(p_-,s_-)+e^+(p_+,s_+)\rightarrow \gamma (k_1,h_\gamma)+Z(k_2,h_Z),
        \label{process}
\end{equation}
where $h_\gamma$ can take values $\pm 1$ and the value for $h_Z$
can be $\pm 1$ and 0.  As in Ref.~\cite{ARSB},
we impose electromagnetic gauge invariance. 
The most general effective $CP$-violating Lagrangian, retaining 
terms upto dimension 6 can be written as~\cite{CR}
\be
\barr{rcl}
{\cal L} &= & \displaystyle
   e \frac{\lambda_1}{ 2 m_Z^2} F_{\mu\nu}
    \left( \p^\mu Z^\lambda \p_\lambda Z^\nu
          - \p^\nu Z^\lambda \p_\lambda Z^\mu
      \right)
       \\[2ex]
& & \displaystyle
      +\frac{e}{16 c_W s_W} \frac{\lambda_2}{m_Z^2}
       F_{\mu\nu}F^{\nu \lambda}
       \left(\p^\mu Z_\lambda + \p_\lambda Z^\mu   \right),
\earr
      \label{lagrangian}
\ee
where $e$ is the electric charge, $m_Z$ is the mass of $Z$ boson,
$c_W=\cos \theta_W$ and $s_W=\sin \theta_W$, with $\theta_W$ as the
weak mixing angle. $\lambda_1$ and $\lambda_2$ are in general 
complex.
Terms involving divergences of the vector fields have been dropped from the
Lagrangian.
\begin{figure}[htb]
\centering
\epsfig{file=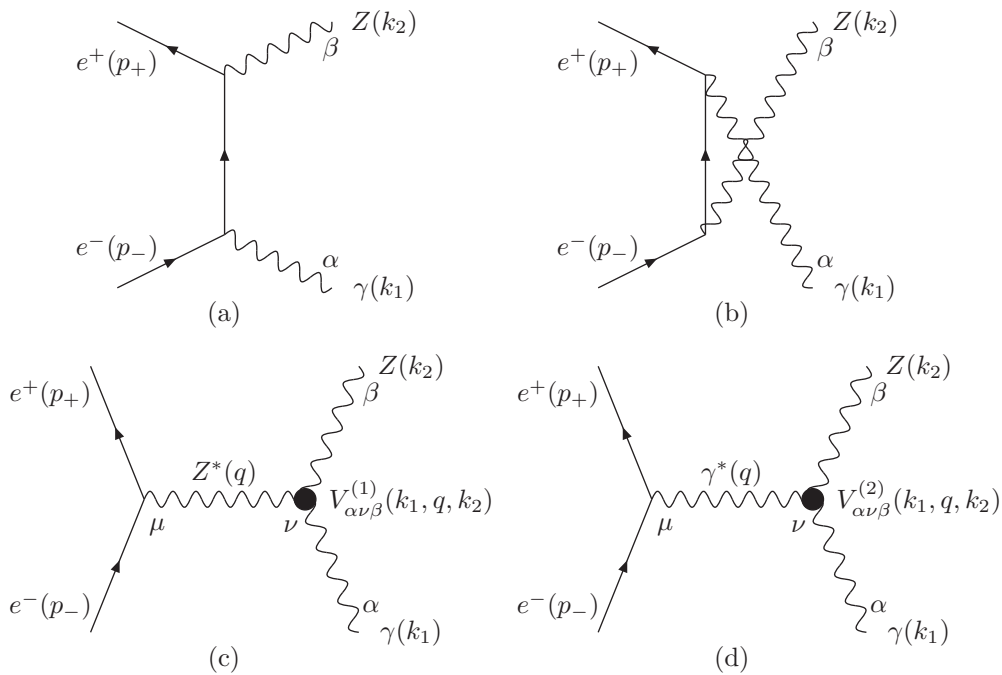}
\caption{Diagrams contributing to the process \eegz. Diagrams (a) and (b)
are SM contributions and diagrams (c) and (d) correspond to contributions from
the anomalous \gzz and \ggz couplings.}
\label{feynman}
\end{figure}
The SM diagrams contributing to the process (\ref{process}) are shown in Figs.
\ref{feynman} (a) and \ref{feynman} (b), which correspond to 
$t$-- and a $u$--channel electron exchange, while
the extra piece in the Lagrangian (\ref{lagrangian}) introduces two
$s$--channel diagrams with $\gamma$-- and $Z$--exchange, respectively, shown in
Figs. \ref{feynman} (c) and \ref{feynman} (d).
Here we have used $q = k_1 + k_2$ as the momentum label for the intermediate 
state in the
$s$ channel, and the tensors $V^{(1)}$ and $V^{(2)}$
corresponding to the three-vector vertices are given by
\be
\barr{rcl}
V^{(1)}_{\alpha\nu\beta}(k_1,q,k_2)
   &= &  \displaystyle
k_1\cdot q \: g_{\alpha\beta} \: k_{2\nu} + k_1\cdot k_2 g_{\alpha\nu} q_\beta
- k_{1\beta} \: q_\alpha \: k_{2\nu}  - k_{1\nu} \:q_\beta \: k_{2\alpha}
   \\[2ex]
V^{(2)}_{\alpha\nu\beta}(k_1,q,k_2)
   & = & \displaystyle
  \frac{1}{2}   \left[
  g_{\alpha\beta}
            \left( k_2\cdot q \: k_{1\nu} - k_1\cdot q \: k_{2\nu} \right)
        - g_{\nu\alpha}
            \left( k_2\cdot q \: k_{1\beta} + k_1 \cdot k_2 \: q_\beta \right)
      \right.
      \\[2ex]
   & &  \displaystyle \hspace{2em}
    \left.
       + g_{\nu\beta}
            \left( k_1\cdot k_2 \: q_\alpha - k_1\cdot q \: k_{2\alpha} \right)
        + q_\alpha \: k_{2\nu} \: k_{1\beta}
        + q_\beta \: k_{1\nu} \: k_{2\alpha}
    \right]
{}.
\earr
     \label{vertices}
\ee
We have computed the cross section incorporating transverse beam polarization
using the helicity amplitudes given in 
~\cite{czyz}. Note that the parametrization of the anomalous
$Z \gamma V^*$ coupling, where $V^*\rightarrow \gamma^*$ or $Z^*$ in~\cite{czyz} is given
in terms of  $f_1^{\gamma}$ and $f_1^Z$.
\footnote{They are related to $\lambda_1$ and 
$\lambda_2$ as, $f_1^\gamma = \frac{\lambda_2}{4}$ and $f_1^Z  =  - \frac{\lambda_1}{4}$.}

The cross section $\Sigma$ for the transversely polarized state 
 can be expressed as \cite{hikasa}
\begin{eqnarray*}
&\displaystyle \Sigma=\Sigma_{\rm unpol}-
{1\over 2} P_T {\overline P_{T}} {\rm Re}\, [T^*_{++} T_{--}]-
{1\over 2} P_T \overline P_{T}  {\rm Re}\, [e^{-2i\phi} T^*_{+-}T_{-+}]
					& \\
& \displaystyle + {1\over 2} P_T {\rm Re} [e^{-i\phi}\left(T^*_{+-} T_{--}+
				       T^*_{++} T_{-+} \right)] -
{1\over 2} {\overline P_T} {\rm Re} [e^{-i\phi}(T^*_{++} T_{-+} 
						   +T^*_{--} T_{-+})], 
						& 
\end{eqnarray*}
where the $T_{++},\, T_{+-},\, T_{-+}$ and $T_{--}$ are helicity
amplitudes for the process at hand, $\phi$ is the final-state azimuthal
angle and $\Sigma_{\rm unpol}$ is the unpolarized cross section. 
In  $T_{ab}$, the subscripts $a,b=+,-$ stand for the helicities
of the $e^+$ and $e^-$ respectively.  In the above, 
beyond standard model (BSM) interactions of the chirality--conserving type contribute to
the amplitudes $T_{+-}$ and $T_{-+}$, while those of the chirality--violating 
type contribute to $T_{aa},a=+,-$ (the SM interactions themselves
contribute only to $T_{+-}$ and $T_{-+}$, when $m_e$ effects are neglected).  
Note also the characteristic $e^{2i\phi}$ dependence accompanying the
terms bilinear in transverse polarization, and the $e^{i\phi}$ dependence
accompanying the linear transverse polarization pieces when the
BSM physics is worked out to leading order.

However the process under study is of the annihilation type and contains
no electron and electron neutrino in final state, so $T_{++}=T_{--}=0$.
So the above expression reduces to a much simplified form which is
given below
\begin{equation}
\displaystyle \Sigma=\Sigma_{\rm unpol}-
{1\over 2} P_T \overline P_{T} (\cos 2\phi {\rm Re}\ T^*_{+-}T_{-+} +
\sin 2\phi ~{\rm Im}~T^*_{+-}T_{-+}).
\end{equation}
This will be used in the following sections to evaluate the
distributions of interest.
 
\section{Distributions in the presence of transverse polarization}\label{distributions}

We now give explicit expression for differential scattering cross
section, for the two cases, viz.,
when the helicity of $Z$ is resolved, summing over the helicity of
$\gamma$, and {\it vice versa}.

Let us introduce the definitions :
\beqar
\sbar & \equiv & \displaystyle \frac{s}{m_Z^2} \nonumber\\
{\cal B} &=&  \frac{\alpha^2}{16 s_W^2 m_W^2 \sbar}
     \left( 1 - \frac{1}{\sbar}   \right)
     (g_V^2+g_A^2),\nonumber\\
\eeqar
where $s$ is the square of the total c. m. energy, and 
$g_V$ and $g_A$ represent the vector and
the axial vector couplings of the electron with $Z$ given by 
\begin{equation}
g_V = -1 + 4\sin^2\theta_W ;\quad g_A = -1
     \label{gVgA}.
\end{equation}

In the equations below only terms of linear order in the anomalous couplings, are retained
since they are
expected to be small. 
So far as the CP-violating part of the differential cross section is
concerned, this is not an approximation, since the only contribution is
from the interference between the SM amplitude and the CP-violating
amplitude, linear in the anomalous couplings. The denominator of the
asymmetries which we calculate, however, would in principle receive a
contribution from terms quadratic in the anomalous couplings. In principle, the 
validity of the linear approximation is dependent on the c. m. energy of the process, and
for  larger  c. m. energy would require the
couplings to be smaller. We will
see that in our case the contribution of the quadratic terms to the
cross section is negligible for  Re $\lambda_1$ and Re $\lambda_2$ 
below about 0.01, where our limits lie, 
and therefore the linear approximation holds good. Expressed differently
it can be said that the observables defined here are not sensitive to
quadratic terms in the cross section, for the values which will be probed at the 
linear collider. 

Resolving the polarization of $Z$ and summing over the polarizations of 
$\gamma$, 
the differential cross sections 
for the production of circularly polarized $Z^{\pm}$ and
longitudinally polarized $Z^L$ are given by
\begin{eqnarray}
{\left. \frac{d\sigma}{d\Omega} \right|}_{Z^{\pm}} &=& 
  {\cal B} \left({\left. \frac{d\sigma_{SM}}{d\Omega}\right|}_{Z^{\pm}}
+ C_A^{Z^{\pm}} \cos \theta + C_B^{Z^{\pm}}  \right)
 \end{eqnarray} 
\begin{eqnarray}
{\left. \frac{d\sigma}{d\Omega} \right|}_{Z^L} &=& 
  {\cal B} \left({\left. \frac{d\sigma_{SM}}{d\Omega}\right|}_{Z^L} +
C_A^{Z^L} \cos \theta + C_B^{Z^L}  \right)
 \end{eqnarray} 
Here
\begin{eqnarray} \label{Z_cross1}
{\left. \frac{d\sigma_{SM}}{d\Omega} \right|}_{Z^{\pm}}& =& \left(1+\frac{2\bar{s}}{(\bar{s}-1)^2}\right)
 \left\lbrace \frac{1}{4 \sin^2 \theta}\left(
   \pm \frac{8 g_A g_V \cos \theta}{g_V^2+g_A^2}+2(1+\cos^2\theta)\right) \right. \nonumber \\ 
 & &\left. - \frac{1}{2} P_e P_{\bar{e}} \frac{g_V^2-g_A^2}{g_V^2+g_A^2}\cos 2 \phi \right\rbrace \nonumber \\ 
   {\left. \frac{d\sigma_{SM}}{d\Omega} \right|}_{Z^L} & =& \frac{2\bar{s}}{(\bar{s}-1)^2}
 \left\lbrace 1 + P_e P_{\bar{e}} \frac{g_V^2-g_A^2}{g_V^2+g_A^2}\cos 2
\phi \right\rbrace  
\end{eqnarray}   
are the corresponding SM differential cross sections, 
and we have defined 
\begin{eqnarray}\label{Z_crosst}
 C_A^{Z^{\pm}} &=& \frac{1}{8(g_V^2+g_A^2)} \left( ((g_A^2-g_V^2)P_e P_{\bar{e}}
 \cos 2 \phi-(g_A^2+g_V^2)) {\rm Im} \lambda_ 1\right. \nonumber \\
  & & \left. +(1+ P_e P_{\bar{e}}\cos 2 \phi) g_V {\rm Im} \lambda_ 2
   + g_A P_e P_{\bar{e}}\sin 2 \phi {\rm Re}\lambda_ 2\right) \nonumber \\
C_B^{Z^{\pm}} &=&  \pm \frac{1}{8(g_V^2+g_A^2)} 
\left( -2 g_A g_V {\rm Im} \lambda_ 1 +(g_A^2-g_V^2)P_e P_{\bar{e}}\sin 2 \phi {\rm Re}\lambda_ 1\right. \nonumber \\
& & +\left.(1+ P_e P_{\bar{e}}\cos 2 \phi) g_A {\rm Im} \lambda_ 2 + g_V P_e P_{\bar{e}}\sin 2 \phi {\rm Re}\lambda_ 2 \right)
 \end{eqnarray} 
 and
 \begin{eqnarray}\label{Z_cross}
 C_A^{Z^L} &=& \frac{-2 \bar{s}}{8(g_V^2+g_A^2)} \left( ((g_A^2-g_V^2)P_e P_{\bar{e}}
 \cos 2 \phi-(g_A^2+g_V^2)) {\rm Im} \lambda_ 1\right. \nonumber \\
  & & \left. +(1+ P_e P_{\bar{e}}\cos 2 \phi) g_V {\rm Im} \lambda_ 2
   + g_A P_e P_{\bar{e}}\sin 2 \phi {\rm Re}\lambda_ 2\right) \nonumber \\
C_B^{Z^L} &=&  0.
 \end{eqnarray} 

Note the appearance of ${\rm Re} \lambda_1$ in $C_B^{Z^{\pm}}$. On
summing over the $Z$ helicities, however, the dependence on ${\rm Re}
\lambda_1$ disappears. In the above, 
as defined earlier, $\theta$  is the angle between photon and the 
beam direction of $e^-$, chosen as the $z$
axis. $\phi$ is the azimuthal angle of the photon with the direction of
the transverse polarization
of the $e^-$ chosen as the $x$ axis. $d\Omega= d\cos\theta d\phi$ is
the integration measure for the angular variables $\theta$ and $\phi$.
The $e^+$ polarization direction can be parallel or antiparallel to
the $e^-$ polarization direction, the polarization in the former case
being taken as positive.

The cross sections for two circular polarization states of $\gamma$ 
summing over all the $Z$ states are given by: 
\begin{eqnarray}
{\left. \frac{d\sigma}{d\Omega} \right|}_{\gamma^\pm} &=& 
  {\cal B} \left({\left. \frac{d\sigma_{SM}}{d\Omega}\right|}_{\gamma^\pm}+ C_A^{\gamma^\pm} \cos \theta + C_B^{\gamma^\pm}  \right)
 \end{eqnarray} 
where
\begin{eqnarray}
 {\left. \frac{d\sigma_{SM}}{d\Omega} \right|}_{\gamma^\pm} &=&  \left[\frac{1}{2 \sin^2\theta}\left(
  1+\cos^2\theta +\frac{4\bar{s}}{(\bar{s}-1)^2}-P_e P_{\bar{e}} \frac{g_V^2-g_A^2}{g_V^2+g_A^2}\sin^2\theta 
  \cos 2\phi
   \right. \right.  \nonumber \\
  & & \left.\left.\mp\frac{4g_A g_V}{g_V^2+g_A^2}\left(\frac{\bar
s+1}{\bar s- 1}\right)\cos\theta \right)  \right]
 \end{eqnarray} 
and
\begin{eqnarray} 
 C_A^{\gamma^\pm} &=& 
 \frac{\bar{s}-1}{8(g_V^2+g_A^2)} \left((g_V^2+g_A^2+(g_V^2-g_A^2)
 P_e P_{\bar{e}} \cos 2 \phi){\rm Im} \lambda_ 1 \right. \nonumber \\
 & & \left. - g_V(1+P_e P_{\bar{e}} \cos 2 \phi ) {\rm Im} \lambda_ 2 - 
 P_e P_{\bar{e}} g_A \sin 2\phi {\rm Re}\lambda_ 2 \right)  \nonumber \\
 C_B^{\gamma^\pm} &=& 
\mp \frac{\bar{s}-1}{8(g_V^2+g_A^2)} 
 \left(2 g_A g_V \left(\frac{\bar s+1}{\bar s-1}\right){\rm Im} \lambda_ 1 
  + P_e P_{\bar{e}} (g_A^2-g_V^2) \sin 2\phi {\rm Re}\lambda_ 1 \right. \nonumber \\
 & & \left. + g_A \left(P_e P_{\bar{e}} \cos 2 \phi-\frac{\bar{s}+1}{\bar{s}-1}\right){\rm Im} \lambda_ 2  +
   g_V P_e P_{\bar{e}} \sin 2 \phi {\rm Re}\lambda_ 2 \right)  .
\end{eqnarray} 
Again, ${\rm Re} \lambda_1$ occurs in  $C_B^{\gamma^\pm}$, but
cancels on summing over the $\gamma$  helicities.

Here one can easily check from Eq. (9) for Z  and Eq. (13) for $\gamma$  
that after summing  over different final helicity states we get the same result, which also agrees with one obtained in\cite{ARSB}
for unpolarized final states.

 At the ILC it is expected that about
90\% electron polarization would be achievable along with a positron polarization of 
60\%~\cite{review}. The beams will be longitudinally polarized,
but there is a possibility that spin rotators can be used
to produce transversely polarized beams.  $P_e$ and $P_{\bar{e}}$ are, respectively, the degrees of
polarization of the $e^-$ and $e^+$; for our calculation we have taken a realistic
value of $P_e$ = 0.8 and $P_{\bar{e}}$ = 0.6. In the next section we will employ these 
expressions to obtain 90\% confidence level(CL) limits on ${\rm Re}\lambda_1$.

\section{Asymmetries and Numerical Results}\label{asymmetries}

In order to make the above expressions useful for
applications at the ILC, and to disentangle the anomalous couplings, we will define
certain asymmetries that will isolate ${\rm Re}\lambda_i$ 
with $i$=1,2.  Since the dependence of the
new couplings  on the laboratory observables such as polar and azimuthal
angles
are different, a suitable choice of asymmetries can help in achieving this goal. For the 
helicity--summed case, it was observed that the contribution from {\rm Re}$\lambda_1$
was zero. But the inclusion of helicity of either of the final-state results in a contribution from
${\rm Re}\lambda_1$.  Therefore the situation where all the spin configurations are available
is explored in this section by constructing various asymmetries.  A through numerical analysis 
is done to put a bound on  the anomalous coupling ${\rm Re}\lambda_1$ along with ${\rm Re}\lambda_2$.

\subsection{Integrated Asymmetries}

We define two CP-odd asymmetries constructed from suitable partial cross
sections. In all case, we assume a cut-off of $\theta_0$ on the polar
angle in the forward
and backward directions, required to stay away from the beam pipe, 
and our asymmetries are therefore functions of
$\theta_0$. The cut-off may be chosen to optimize the sensitivity. 
One of the asymmetries,
$ A^V_\pm(\theta_0)$, combines a forward-backward asymmetry along with an azimuthal symmetry.
The other, $ A'^V_\pm(\theta_0)$, is an asymmetry only in $\phi$. The
asymmetries are given by
\begin{eqnarray}\label{asym1}
& \displaystyle  A^V_\pm(\theta_0)=
{1\over \sigma ^V_{SM}(\theta_0)}
\sum_{n=0}^3 (-1)^n
\left(
\int_{0}^{\cos \theta_0} d \cos\theta -
\int_{-\cos \theta_0}^{0} d \cos\theta \right) 
 \int_{\pi n/ 2}^{\pi(n+1)/  2} d\phi \,
{{\left. d \sigma'_\pm \over d \Omega \right |}_V}   
\end{eqnarray}
\begin{eqnarray}\label{asym2}
& \displaystyle A'^V_\pm(\theta_0)=
{1\over \sigma ^V_{SM}(\theta_0)}
\sum_{n=0}^3 (-1)^n
\int_{-\cos \theta_0}^{\cos \theta_0} 
\left( \int_{\pi n/ 2}^{\pi(n+1)/  2} d\phi \,
{{\left. d \sigma'_\pm \over d \Omega \right|}_V}  \right) d \cos\theta    
\end{eqnarray}
with the SM cross section given by
\begin{eqnarray}
& \displaystyle  \sigma ^V_{SM}(\theta_0)=
\int_{-\cos \theta_0}^{\cos \theta_0} d \cos\theta
\int_{0}^{2 \pi} d\phi \,
{{\left. d \sigma_{SM} \over d \Omega  \right|}_V  }\,  &
\end{eqnarray}
where V can be $\gamma$ or $Z$ depending on whose polarization is being considered.  Since 
we are mainly concentrating on the coupling {\rm Re}$\lambda_1$, from Eq.~\ref{Z_cross} we
see that longitudinal $Z$ with $h=0$ is not sensitive to this coupling. Therefore we will be mainly
concentrating on  $Z^h$ with $h=\pm$.  

Here $A^V_\pm(\theta_0)$ and $A'^V_\pm(\theta_0)$ are calculated for different combinations
of  ${\left. \frac{d \sigma'_\pm}{d \Omega}\right|}_V$ which is the value obtained 
when the two helicity states of $V$ are summed over or the difference is taken.
\begin{equation}\label{hel_ind}
{\left. \frac{d \sigma'_+}{d \Omega}\right|}_V = {\left. \frac{d
\sigma}{d \Omega}\right|}_{V^+}
+  {\left. \frac{d \sigma}{d \Omega}\right|}_{V^-}
\end{equation}

\begin{equation}\label{hel_dep}
{\left.\frac{d \sigma'_-}{d \Omega}\right|}_V = {\left. \frac{d \sigma}{d \Omega}\right|}_{V^+}
-  {\left. \frac{d \sigma}{d \Omega}\right|}_{V^-}.
\end{equation}
Here $V^{\pm}$ refers to $V$ with helicity $\pm 1$. The above choice of asymmetries
is motivated by the purpose of isolating the couplings {\rm Re}$\lambda_1$
and  {\rm Re}$\lambda_2$. Equation~(\ref{hel_ind}) contains a term proportional to
$\cos\theta$ coming from $C^{V^\pm}_A$, which with polar angle forward-backward asymmetry
in  $A^V_\pm(\theta_0)$ survives, whereas $A'^V_\pm(\theta_0)$ goes to zero.
Similarly Eq.~(\ref{hel_dep}) contains term independent of $\cos\theta$ coming 
from $C^{V^\pm}_B$, therefore the polar angle
forward-backward asymmetry of $A^V_\pm(\theta_0)$ is equal to zero, and
$A'^V_\pm(\theta_0)$ survives. In this case the term proportional to $\cos\theta$
coming from ${\left. \frac{d\sigma_{SM}}{d\Omega} \right|}_{V^{\pm}}$ vanishes for both
$A^V_\pm(\theta_0)$ and $A'^V_\pm(\theta_0)$.

\begin{enumerate}
\item
Considering the case when the helicity of $Z$ is kept and that of $\gamma$ is
summed over:

The asymmetries from Eq.~(\ref{hel_ind}) now evaluate to :
\begin{eqnarray}\label{Z_dep1}
A^Z_+(\theta_0) &=& \frac{{\cal B}g_A P_e P_{\bar{e}} \cos^2\theta_0 {\rm Re}\lambda_2}
     {(g_V^2+g_A^2)\sigma ^Z_{SM}(\theta_0)} \nonumber \\
A'^Z_+(\theta_0)&=& 0
\end{eqnarray}
Similarly from Eq.~(\ref{hel_dep}) the asymmetries are :
\begin{eqnarray}\label{Z_dep2}
A^Z_-(\theta_0) &=& 0 \nonumber  \\
A'^Z_-(\theta_0) &=& \frac{2{\cal B} P_e P_{\bar{e}} ((g_A^2-g_V^2) {\rm Re}\lambda_1 + g_V {\rm Re}\lambda_2) \cos \theta_0}
     {(g_V^2+g_A^2)\sigma ^Z_{SM}(\theta_0)}
\end{eqnarray}
where 
\begin{equation}
\sigma ^Z_{SM}(\theta_0) =4 \pi {\cal B}\frac{\bar{s}^2+1}{(\bar{s}-1)^2}\left[
   {\rm ln}\left(\frac{1+\cos
\theta_0}{1-\cos\theta_0}\right)-\cos\theta_0 \right]  .
\end{equation}
In Eq.~(\ref{Z_dep2}) since $g_V$ is small compared to $g_A^2 - g_V^2$,
the asymmetry will be dominated by the {\rm Re}$\lambda_1$ term.
We here therefore expect to obtain a more stringent limit on Re$\lambda_1$ 
compared to that on Re$\lambda_2$.

\item
Now considering the case that helicity of $\gamma$ is measured 
with that of $Z$ 
summed over:

The asymmetries from Eq.~(\ref{hel_ind}) are :
\begin{eqnarray}\label{G_dep1}
A^\gamma_+(\theta_0) &=& - \frac{{\cal B}g_A P_e P_{\bar{e}} (-1+\bar{s})\cos^2\theta_0 {\rm Re}\lambda_2}
     {(g_V^2+g_A^2)\sigma ^\gamma_{SM}(\theta_0)} \nonumber \\
A'^\gamma_+(\theta_0)&=& 0.
\end{eqnarray}
Here from Eq.~(\ref{hel_dep}) the asymmetries are :
\begin{eqnarray}\label{G_dep2}
A^\gamma_-(\theta_0) &=&   0 \nonumber  \\
A'^\gamma_-(\theta_0) &=& -\frac{2{\cal B} P_e P_{\bar{e}} ((g_A^2-g_V^2) {\rm Re}\lambda_1 + g_V {\rm Re}\lambda_2)(-1+\bar{s}) \cos \theta_0}
     {(g_V^2+g_A^2)\sigma ^\gamma_{SM}(\theta_0)},
\end{eqnarray}
where 
\begin{equation}
\sigma ^\gamma_{SM}(\theta_0) =4 \pi {\cal B}\left[\frac{\bar{s}^2+1}{(\bar{s}-1)^2}
   {\rm ln} \left(\frac{1+\cos
\theta_0}{1-\cos\theta_0}\right)-\cos\theta_0 \right]  .
\end{equation}
As before, since $g_V$ is small compared to $g_A^2 - g_V^2$,
we here expect to obtain more stringent limit
on Re$\lambda_1$ compared to that on Re$\lambda_2$. 
\end{enumerate}

Analyzing  Eqs.~(\ref{Z_dep1}), (\ref{Z_dep2}), (\ref{G_dep1}) and (\ref{G_dep2}), it is seen that 
for both $Z$ and $\gamma$,  $A^V_+(\theta_0)$ is only sensitive to {\rm Re}$\lambda_2$, whereas $A'^V_-(\theta_0)$
depends on both {\rm Re}$\lambda_i$, $i=1,2$, where $V$ = $Z$, $\gamma$. The above is due to the 
polar angle dependent term in Eq.~(\ref{hel_ind}) 
surviving for $A^V_+(\theta_0)$, and the helicity dependent
polar angle independent term in Eq.~(\ref{hel_dep}) surviving for $A'^V_-(\theta_0)$.  Comparing
our results to the earlier work where all the three helicity states of $Z$ are summed over~\cite{ARSB},
the same asymmetry  $A^V_+(\theta_0)$ evaluated for Eq.~(\ref{hel_ind}), for the two transverse states of $Z$
is smaller by a factor of $(\bar{s}-1)$.  Because of this the limit on {\rm Re}$\lambda_2$
is poorer in this case, whereas  the additional advantage here is that we can put a limit on 
{\rm Re}$\lambda_1$ which was not possible earlier.  The analysis done for the case of $\gamma$
does not encounter this problem as here we are summing over all its helicity states, 
unlike the case of $Z$.

\subsection{Numerical Analysis} 

We have calculated the cross section and the asymmetries for the case when $e^+$ polarization
is parallel to 
$e^-$.  For our sensitivity analysis,
we have assumed an integrated luminosity
of 500 ${\rm fb}^{-1}$. 

\begin{figure}[!]
\begin{minipage}[!]{0.45\linewidth}
\centering
\includegraphics[scale=0.7]{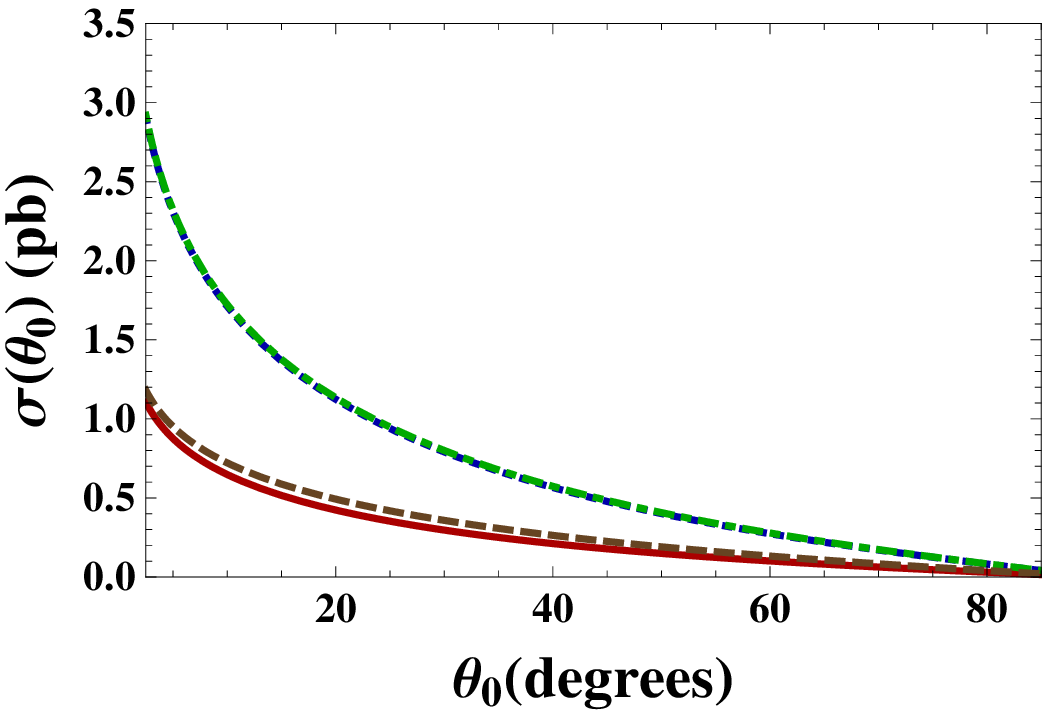}
\caption{{\small The total cross section $\sigma(\theta_0)$ plotted
as a function of  $\theta_0$ for 
({\rm Re}$\lambda_2$ = 0, {\rm Im}$\lambda_1$ = 0, {\rm Im}$\lambda_2$ = 0)
at $\sqrt{s}$ = 800 GeV, (SM, red-solid) 
({\rm Re}$\lambda_1$=0.01, brown-dashed) 
and $\sqrt{s}$ = 500 GeV (SM, blue-dotted) ({\rm Re}$\lambda_1$=0.01, green-dot-dashed). 
green-dot-dashed is coincident with blue-dotted}}
\label{fig.sm}
\end{minipage}
\hspace{0.5cm}
\begin{minipage}[!]{0.45\linewidth}
\centering
\includegraphics[scale=0.7]{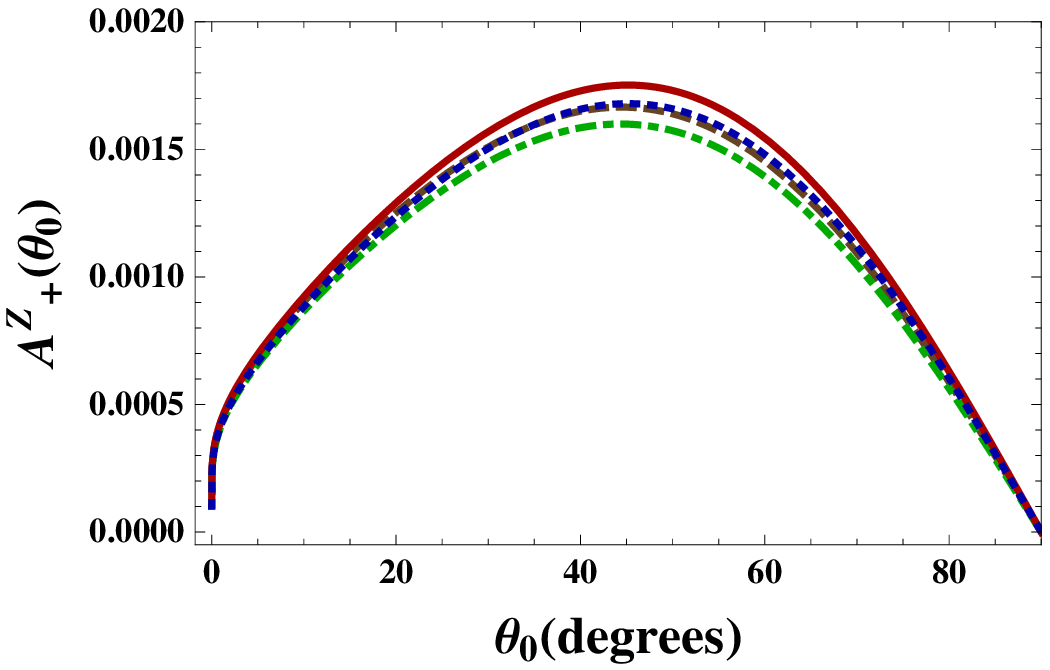}
\caption{{\small The asymmetry $A^Z_+(\theta_0)$ plotted as a function of cut-off $\theta_0$
for a value of {\rm Re}$\lambda _2$=0.1 at $\sqrt{s}$= 800 GeV  with (brown-dashed) 
and without (red-solid) quadratic terms 
and 500 GeV with (green-dot-dashed) and without (blue-dotted) quadratic terms. Brown-dashed 
is coincident with blue-dotted}}
\label{fig:relamZ}
\end{minipage}
\end{figure}

Figure~\ref{fig.sm} shows the total cross section $\sigma(\theta_0)$, plotted as a function of cut off angle $\theta_0$
at $\sqrt{s}$ = 500 GeV and 800 GeV. The anomalous couplings 
contribute to the total cross section only at  quadratic order~\cite{CR}. We
have kept these terms in Fig.~\ref{fig.sm}, for a particular combination.
Similar behavior occurs for the case $\lambda_2=0.01$ and $|\lambda_1| = 0$, since the cross section is
almost symmetric in  $\vert \lambda_1 \vert$ and  $\vert \lambda_2 \vert$
with $(g_V^2 +g_A^2)$ almost equal to 1. It is clear from the figure that
the contribution of anomalous couplings is negligible for values
below 0.01, which correspond to the limits we find from asymmetries given later.
The corresponding limits on Re$\lambda_1$ are 
0.0048 at $\sqrt{s} =500$ Gev and 0.0014 at $\sqrt{s} =800$ Gev for cut off angle
$\theta_0$ = $25\,^{\circ}$. The same limits will hold for Re$\lambda_2$ piece.
Later in this section, we will see that these limits are more stringent than the  $Z$ case
but comparable to the case when the helicity of $\gamma$ is considered.
The total cross section receives contribution both from CP--conserving and CP--violating
anomalous couplings. Therefore, measuring deviation of the total cross section from the SM prediction
would not really be a test for CP--violating couplings which we are considering here.
The linear contribution from the real part of
the CP--violating couplings only occur in the case of transverse polarization 
and has no effect on the total cross section. 
The imaginary parts contribute in linear order
to the differential cross section, but their effect is washed out 
in the total cross section after the $\theta$ integration. 
      
      We have done our whole numerical analyses for two different cases at $\sqrt{s}$
of 500 and 800 GeV. In the first case, we drop the terms quadratic in anomalous couplings
from the denominator of the asymmetries Eqs.~(\ref{asym1}), (\ref{asym2}). 
The best limits on the couplings obtained in this case
are of the order of $10^{-3}$. For the second case, we then include the
terms quadratic in anomalous couplings to the denominator of 
Eqs.~(\ref{asym1}), (\ref{asym2})  
with the value coming from the best limit obtained in the earlier case.
It is apparent from the above discussion that, this contribution of order $10^{-3}$
in the denominator would not much affect the asymmetries and the limits obtained from them.
This is another way of putting the argument that the asymmetries and the limits obtained
are not sensitive to the quadratic terms  in the denominator  at the level in which 
they can be probed. We present the results in detail for the first case, i.e.,
without including quadratic terms, and then show the effect on inclusion of quadratic terms.

\begin{figure}[!]
\begin{minipage}[!]{0.45\linewidth}
\centering
\includegraphics[scale=0.7]{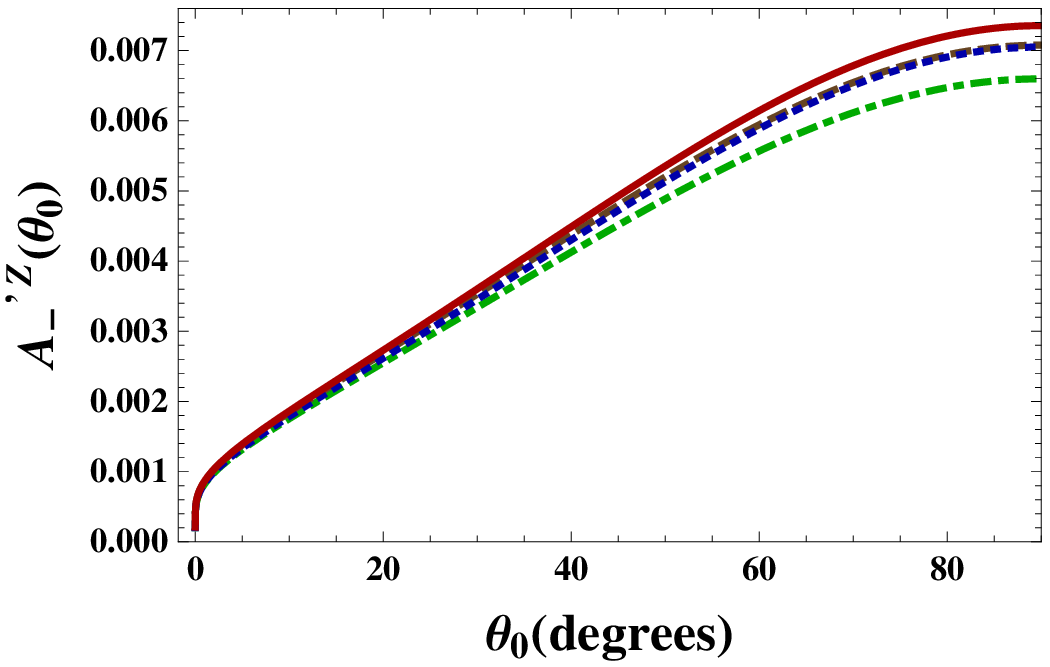}
\caption{{\small The asymmetry $A'^Z_-(\theta_0)$ plotted as a function of cut-off $\theta_0$
for a value of {\rm Re}$\lambda_1$=0.1,\,{\rm Re}$\lambda_2$=0 for different center of mass energies
of 800 GeV with (brown-dashed) 
and without (red-solid) quadratic terms 
and 500 GeV with (green-dot-dashed) and without (blue-dotted) quadratic terms. Brown-dashed 
is coincident with blue-dotted}}
\label{fig:relamZ1}
\end{minipage}
\hspace{0.5cm}
\begin{minipage}[!]{0.45\linewidth}
\centering
\includegraphics[scale=0.7]{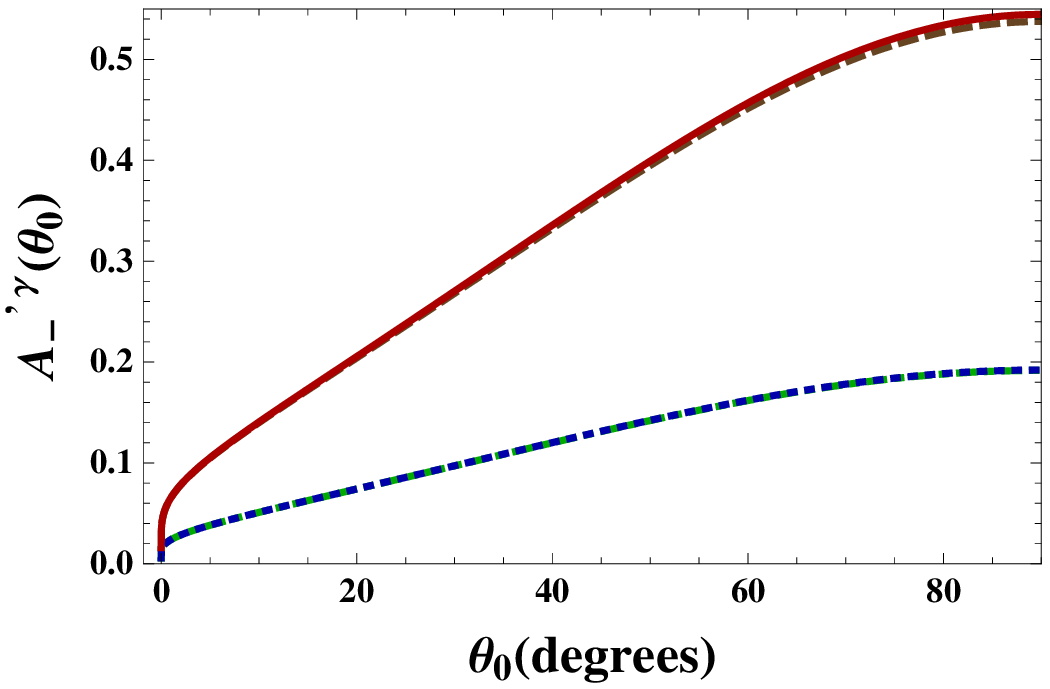}
\caption{{\small The asymmetry $A'^\gamma_-(\theta_0)$ plotted as a function of cut-off $\theta_0$
for a value of {\rm Re}$\lambda_1$=0.1,\,{\rm Re}$\lambda_2$=0 for different center of mass energies
of 800 GeV with (brown-dashed) 
and without (red-solid) quadratic terms 
and 500 GeV with (green-dot-dashed) and without (blue-dotted) quadratic terms. Green-dot-dashed
is coincident with blue-dotted}}
\label{fig:relamG1}
\end{minipage}
\end{figure}

We now consider the case when the helicity of $Z$ is resolved.
Figure~\ref{fig:relamZ} shows the $\theta$ integrated version of the asymmetry $A^Z_+(\theta_0)$ 
plotted as a function of cut-off $\theta_0$ for different c.m. energies
with a value of {\rm Re}$\lambda_2$=0.1.
Figure~\ref{fig:relamZ1} shows $A'^Z_-(\theta_0)$ as a function of cut-off $\theta_0$,
when the value of the anomalous couplings are taken as {\rm Re}$\lambda_1$ =0.1
and {\rm Re}$\lambda_2$ =0 for different c.m. energies.
However, as mentioned before, {\rm Re}$\lambda_2$, in the case of $A'^Z_-(\theta_0)$,
is accompanied
by the numerically small coefficient and thus we obtain a larger limit on Re$\lambda_2$
where our linear approximation is not valid. So we will drop Re$\lambda_2$ in our 
consideration of $A'^Z_-(\theta_0)$ from here onwards. $A^Z_+(\theta_0)$ behaves differently
from $A'^Z_-(\theta_0)$ due to the presence
of the $\cos^2\theta_0$ term in the numerator. The asymmetry in the
latter case increases with the cut off due to the SM cross section in the denominator which decreases 
faster than the numerator. In all the above cases the asymmetry does not 
change much with c. m. energy as the observables above are all independent of it.
We have not given the figure corresponding to $A^\gamma_+(\theta_0)$ since
it has already been considered in Ref.~\cite{ARSB}. The figures are all plotted
with and without including quadratic terms in the denominator of Eq.~(\ref{asym1}), (\ref{asym2}).
It is observed that there is not much deviation, with the inclusion of quadratic terms
couplings for the values, at the level probed by the linear collider.

A similar analysis follows for the case when 
the helicity of $\gamma$ is resolved while that of $Z$ is summed over.  
Figure~\ref{fig:relamG1} shows the 
$\theta$ integrated version of $A'^\gamma_-(\theta_0)$ as a function of cut-off $\theta_0$, 
when the values of the anomalous
couplings are taken as {\rm Re}$\lambda_1$ =0.1
and {\rm Re}$\lambda_2$ =0 for different c.m. energies.  The behavior is the same as that of the case when 
helicity of $Z$ is resolved
except the fact that here the asymmetries are enhanced by an extra factor of about 30 at $\sqrt{s}$ =500 GeV
and a factor of about 75 at  $\sqrt{s}$ =800 GeV
due to the presence of the term $(\bar{s}-1)$ in the numerator of Eq.~(\ref{G_dep2}). Because of this boost  factor we get limit on Re$\lambda_2$ which is well under the ambit of linear approximation and thus
we will retain it for further consideration in $A'^\gamma_-(\theta_0)$. Moreover
due to the presence of this enhancement term, the asymmetry $A'^\gamma_-(\theta_0)$
is sensitive to c.m. energy. In this case also the asymmetries and the limits are calculated with and without
including quadratic terms in the denominator.


\begin{figure}[!]
\begin{center}
\includegraphics[width=6 cm,height=6cm]{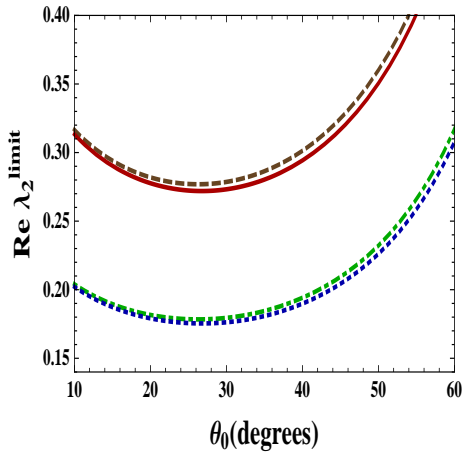}
\caption{{\small 90 \% CL limit possible on {\rm Re}$\lambda_2$ from $A^Z_+(\theta_0)$
of Eq.~(\ref{Z_dep1}) with an integrated luminosity of 500 ${\rm fb}^{-1}$ at
$\sqrt{s}$ = 800 GeV with (brown-dashed) 
and without (red-solid) quadratic terms 
and 500 GeV with (green-dot-dashed) and without (blue-dotted) quadratic terms
plotted as a function of $\theta_0$ }}
\label{fig:relam2lim}
\end{center}
\end{figure}

The asymmetries are used to calculate 90\% CL limits 
with realistic integrated luminosities in the absence of any signal at ILC.
The limit on the coupling is related to
the value $A$ of the asymmetry by:
\beq
\lambda^{\rm lim} = \frac{1.64}{|A|\sqrt{N_{SM}}},
\label{lim_1}
\eeq
where $N_{SM}$ is the number of SM events and A is the value of the asymmetry
for unit value of the coupling.  
The coefficient 1.64 may be obtained
from statistical tables for hypothesis testing with one estimator;
see, e.g., Table 33.1 in Ref.~\cite{RPP}.

\begin{figure}[!]
\begin{minipage}[b]{0.45\linewidth}
\centering
\includegraphics[scale=0.7]{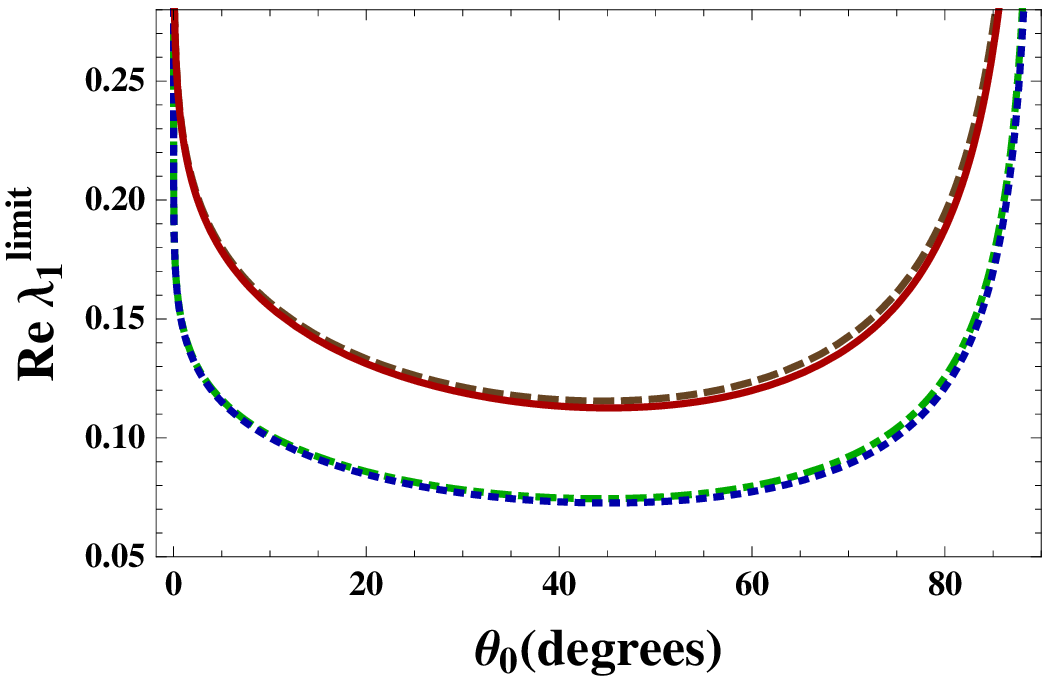}
\caption{{\small 90 \% CL limit on {\rm Re}$\lambda_1$ from $A'^Z_-(\theta_0)$
 with an integrated luminosity of 500 ${\rm fb}^{-1}$ at $\sqrt{s}$ = 800 GeV with (brown-dashed) 
and without (red-solid) quadratic terms 
and 500 GeV with (green-dot-dashed) and without (blue-dotted) quadratic terms
plotted as a function of $\theta_0$. Green-dot-dashed is coincident with blue-dotted}}
\label{fig:relam1Zlim}
\end{minipage}
\hspace{0.5cm}
\begin{minipage}[b]{0.45\linewidth}
\centering
\includegraphics[scale=0.7]{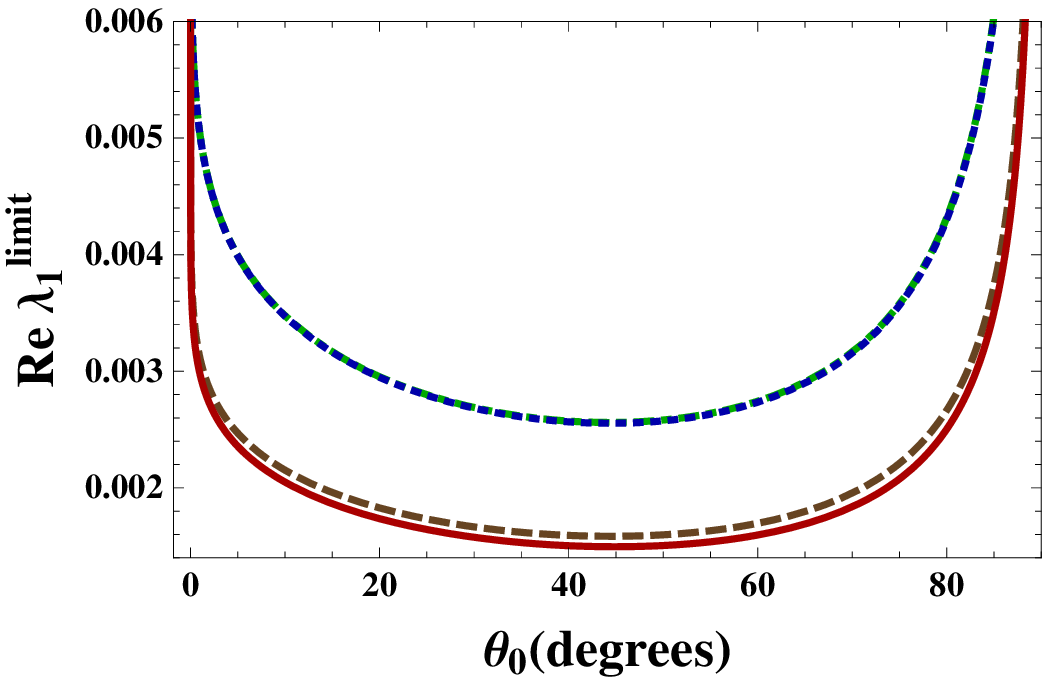}
\caption{{\small 90 \% CL limit on {\rm Re}$\lambda_1$ from $A'^\gamma_-(\theta_0)$
 with an integrated luminosity of 500 ${\rm fb}^{-1}$ at $\sqrt{s}$ = 800 GeV with (brown-dashed) 
and without (red-solid) quadratic terms 
and 500 GeV with (green-dot-dashed) and without (blue-dotted) quadratic terms
plotted as a function of $\theta_0$. Green-dot-dashed is coincident with blue-dotted}}
\label{fig:relam1Glim}
\end{minipage}
\end{figure}

We see from Eqs.~(\ref{Z_dep1}) and (\ref{G_dep1}) that $A^V_+(\theta_0)$  solely depends
on {\rm Re}$\lambda_2$, therefore an independent limit can be placed on it.
Considering the helicity of $Z$, Fig.~\ref{fig:relam2lim} shows
that the best limit for {\rm Re}$\lambda_2$ from $A^Z_+(\theta_0)$
is obtained for $\theta_0$ = $25\,^{\circ}$ at $\sqrt{s}$ = 500 GeV, though any nearby values of
$\theta_0$ will give the same results.  The limit corresponding to {\rm Re}$\lambda_2$
is 0.1757. A poor limit is obtained for $\sqrt{s}$ = 800 GeV,
as the SM cross section in the denominator of Eq.~(\ref{lim_1}) decreases much faster with the
c.m. energy compared to the asymmetry, which does not change. This can be
understood much more clearly from Eqs.~(\ref{Z_dep1}) and (\ref{Z_dep2}), where the anomalous couplings
are not sensitive to the c.m. energies for the transversely polarized $Z$.
Similar analysis is carried out for $A'^Z_-(\theta_0)$. Since $A'^Z_-(\theta_0)$ depends on 
{\rm Re}$\lambda_1$,  Fig.~\ref{fig:relam1Zlim}  shows that the best limit on {\rm Re}$\lambda_1$ is obtained for
$\theta_0$ = $50\,^{\circ}$ at  $\sqrt{s}$ = 500 GeV. The limit on
for {\rm Re}$\lambda_1$ is 0.0958. The limits obtained do not change much with the inclusion
of quadratic terms. 

The results obtained on repeating the earlier analysis for the case when the 
helicity of the $\gamma$ is resolved,  for different c. m. energies, 
is shown in Figure~\ref{fig:relam1Glim}. In this case the sensitivity improves with 
c. m. energy. This is in contrast to the case where the helicity of $Z$
is resolved,
as from Eq.~(\ref{lim_1}), the SM cross section, which decreases with the
c.m. energy, is compensated by the asymmetry $|A|$, which increases much more rapidly 
with energy, due to the term $(\bar{s}-1)$. This also explains the better limits obtained 
on the couplings in this case.
The best limit is obtained for $\theta_0$ = $50\,^{\circ}$ with
{\rm Re}$\lambda_1 $ = 0.0033 at $\sqrt{s}$ = 500 GeV whereas
for $\sqrt{s}$ = 800 GeV, the limits improve to
{\rm Re}$\lambda_1$ = 0.0020 for $\theta_0$ = $45\,^{\circ}$.


We have also evaluated the simultaneous 90\% CL limits that can be obtained on
{\rm Re}$\lambda_2$ and {\rm Re}$\lambda_1$ from $A^V_+(\theta_0)$ and $A'^V_-(\theta_0)$
for the case $V=\gamma$.  In case of $\gamma$, the numerator in Eq.~\ref{G_dep2} is accompanied 
by a factor $(-1+\bar{s})$, resulting in an enhancement. 
The region enclosed by the 
contours obtained by equating the asymmetry with
{\rm Re}$\lambda_2$ as well as {\rm Re}$\lambda_1$ simultaneously
nonzero to 
$\frac{2.15}{\sqrt{N_{SM}}}$
corresponds to the region allowed at the 90 \% CL.
The coefficient 2.15 may be obtained
from statistical tables for hypothesis testing with two estimators,
see, e.g., Table 33.2 in Ref.~\cite{RPP}.
\begin{figure}[!htb]
\begin{center}
\includegraphics[width=6 cm,height=6cm]{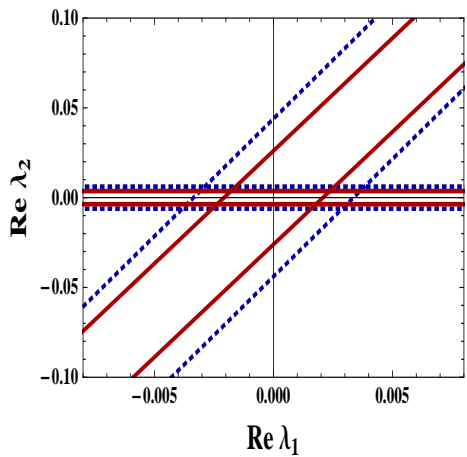}
\caption{{\small 90 \% CL contours for the simultaneous determination of {\rm Re}$\lambda_2$ 
and {\rm Re}$\lambda_1$  from $A^\gamma_+(\theta_0)$  and  $A'^\gamma_-(\theta_0)$
 with an integrated luminosity of 500 ${\rm fb^{-1}}$ at $\sqrt{s}$ = 800 GeV 
(red-solid) and $\sqrt{s}$ = 500 GeV (blue-dotted) by including quadratic corrections.
The area inside the parallelogram is the allowed region.  }}
\label{fig:simGlim}
\end{center}
\end{figure}
The above equation is solved for the value of $\theta_0$ giving the best limit
for the couplings, which for $A^\gamma_+(\theta_0)$ is $25\,^{\circ}$, and similarly for
$A'^\gamma_-(\theta_0)$ we take $\theta_0$ as $45\,^{\circ}$  at $\sqrt{s}$ = 500 GeV.
Whereas at $\sqrt{s}$ = 800 GeV, the best limits for the couplings are obtained at
$27\,^{\circ}$ for $A^\gamma_+(\theta_0)$ and $45\,^{\circ}$ for $A'^\gamma_-(\theta_0)$.
Figure~\ref{fig:simGlim} shows the simultaneous limit
obtained on {\rm Re}$\lambda_i$ taking the helicity of $\gamma$, for $\sqrt{s}$ = 500 GeV and 800 GeV. The 
individual limits obtained by taking one coupling to be nonzero at a time as well as
the simultaneous limit on the anomalous couplings from the asymmetries 
for different c.m. energies are shown in 
Tables. \ref{G_table1} and \ref{G_table2}.  
It is seen that a
better individual limit is obtained on {\rm Re}$\lambda_1$, compared to the
simultaneous limit.
On the other hand, the simultaneous limit on {\rm Re}$\lambda_2$ is better than
than the individual limit obtained on it from the asymmetry $A'^\gamma_-(\theta_0)$. 
This is due to the fact that the coefficient $g_V$ accompanying {\rm Re}$\lambda_2$
is too small, Eq.~(\ref{G_dep2}),  to give a deviation from standard model results. But
$A^\gamma_+(\theta_0)$ gives a better individual limit on {\rm Re}$\lambda_2$ compared to the simultaneous
case as here the accompanying term is $g_A$ much larger compared to $g_V$, Eq.~(\ref{G_dep1}).

%
%
%
\begin{table}[!htb]
\centering
\begin{tabular}{|l|c|c|c|c}
\hline
Coupling & \multicolumn{2}{|c|} {Individual limit from}
 & Simultaneous limits \\ \cline{2-3}
& $A^\gamma_+(\theta_0)$ & $A'^\gamma_-(\theta_0)$  &\\ 
\hline 
Re $\lambda_1$ &  &$3.36\times 10^{-3}$  & $3.83\times 10^{-3}$\\
Re $\lambda_2$ & $6.14\times 10^{-3}$ & $4.39\times 10^{-2}$&$ 6.14\times 10^{-3}$  \\
\hline
\end{tabular}
\caption{90 \% CL limits on the couplings from asymmetries $A^\gamma_+(\theta_0)$ for a cut-off
angle of 25$^{\circ}$, and $A'^\gamma_-(\theta_0)$ for a cut-off
angle of 45$^{\circ}$  at $\sqrt{s} = 500$ GeV, and integrated luminosity of 500
${\rm fb}^{-1}$ when the helicity of $\gamma$ is considered with the helicity of
$Z$ summed over, by including quadratic coupling terms in the cross section }
\label{G_table1}
\end{table} 
\begin{table}[!htb]
\centering
\begin{tabular}{|l|c|c|c|c}
\hline
Coupling & \multicolumn{2}{|c|} {Individual limit from}
 & Simultaneous limits \\ \cline{2-3}
& $A^\gamma_+(\theta_0)$ & $A'^\gamma_-(\theta_0)$  &\\ 
\hline 
Re $\lambda_1$ &  &$2.08\times 10^{-3}$  & $2.37\times 10^{-3}$\\
Re $\lambda_2$ & $3.65\times 10^{-3}$ & $2.61\times 10^{-2}$&$ 3.65\times 10^{-3}$  \\
\hline
\end{tabular}
\caption{90 \% CL limits on the couplings from asymmetries $A^\gamma_+(\theta_0)$ for a cut-off
angle of 27$^{\circ}$, and $A'^\gamma_-(\theta_0)$ for a cut-off
angle of 45$^{\circ}$  at $\sqrt{s} = 800$ GeV, and integrated luminosity of 500
${\rm fb}^{-1}$ when the helicity of $\gamma$ is considered with the helicity of
$Z$ summed over, by including quadratic coupling terms in the cross section}
\label{G_table2}
\end{table}

\section{Utilizing Final-State $Z$ and $\gamma$ Helicities}\label{discussion}

In this section we discuss how the final-state helicities of the $Z$ and
the $\gamma$ can be utilized in practice.  We begin by discussing the
case of the helicity of the $Z$.  The discussion closely parallels
the discussion that was recently provided for measuring the helicity
of the top-quark in $t\bar{t}$ production given in Ref.~\cite{APR}. 
So far we have assumed that it would be possible to
isolate a sample of events where the $Z$ has a definite
helicity, which in practice is not possible as one can only
measure polarization at a statistical level. Unlike an incoming beam of
particles, which can be prepared in a pure spin state, an outgoing
particle is not available in a pure state, but only a mixed state,
yielding only an average polarization. In order to be able to make use
of the definitions of various asymmetries which we discuss, we propose
a practical method which would serve to provide a sample with
predominantly positive or negative Z helicities, 
which would lead to a depletion of the efficiency, but would be able to 
achieve the main objective.

The spin of an unstable particle like the $Z$ can be analyzed by looking
at the distribution of its decay products. The decay
distribution of a lepton produced from a $Z$ with a definite helicity  
in the rest frame of the $Z$ is given by
\be\label{polT}
\frac{1}{\Gamma(Z \to \ell^+\ell^-)}\frac{d\Gamma(Z^{\pm}\to\ell^+\ell^-)}
{d\cos\theta_\ell} = \frac{3}{8} \left[ 1 + \cos^2 \theta_\ell \mp
\frac{2g_Vg_A}{g_V^2 + g_A^2} \cos\theta_\ell\right],
\ee
\be\label{pol0}
\frac{1}{\Gamma(Z \to \ell^+\ell^-)}\frac{d\Gamma(Z^L\to\ell^+\ell^-)}
{d\cos\theta_\ell} = \frac{3}{4} \sin^2 \theta_\ell,
\ee
where $\theta_\ell$ is the angle made by the momentum of the $\ell^+$
with the spin quantization axis of the $Z$. The quantity which
differentiates between the positive and negative helicity $Z$
distributions is
\be
\frac{2g_Vg_A}{g_V^2 + g_A^2} \approx  0.147.
\ee
For an elementary derivation of this result, see Sec. 10.2 in
ref.~\cite{Renton}.

It is thus seen from Eqs. (\ref{polT}), (\ref{pol0}) that the $\ell^+$ tends to
be emitted dominantly in the backward direction relative to 
the spin of the $Z$ for
transverse polarization, and peaks at $\theta_\ell = \pi/2$ for
longitudinal polarization. Thus, by
applying a cut keeping dominant emission of the fermions in the
direction of the boost of the $Z$, one can obtain a sample which is
enriched in events with negative helicity. Similarly, 
a cut keeping dominantly backward emission will yield a sample enriched
in events with positive helicity. Finally, keeping leptons emitted
mostly in the transverse direction will give a sample with dominantly
zero helicity.

Thus, one has to actually generate events
including $Z$ decay and use the formulas to make predictions, and then
compare the expected number of events for a given set of anomalous
couplings with experiment and then place a limit.
Such a procedure would give limits which are less stringent than
obtained in our analysis.
Strictly speaking, we should include full spin density matrices for 
$Z$ production  as well as decay into a certain final
state, and consider asymmetries constructed out of the momenta of the
decay products. However, we expect that the procedure described here
will approximate such a complete description, with some reduction of
efficiency.

To get some quantitative idea of the efficiency, we note that if we use
the three charged-lepton channels for $Z$ decay, with a combined branching 
ratio of about 10\%, with the simplifying
assumption that the $\tau$-pair detection efficiency is 1, the
sensitivity is a factor of $\sqrt{10} \approx 3$ of less. The inclusion
of a $b\bar b$ channel would improve the sensitivity somewhat.
A full analysis including $Z$ decay entails a more complicated analysis 
with a different final state, and is beyond the scope of this work.

For projecting the final-state $\gamma$ helicity, there
is no analogous method, as  it is stable. In the context of  the 
Goldhaber neutrino helicity experiment \cite{Goldhaber},
photons of a particular helicity were filtered out by the means of a  
magnetic material.
Depending on the helicity state, photons are either absorbed 
in the material or not. 
Thus, by counting the events in the photopeaks observed with the 
scintillation detector for 
two different polarizations of the magnet, they determined  the photon helicity.
Here also one can 
conceive of using such a material in the construction
of the ILC detector which could be used for polarization determination. We note 
here that, in the context of photon polarimetry measurements, it
has been demonstrated by the CERN NA59 collaboration\cite{Na59Collb} that an aligned-crystal 
technology can be used for an
accurate measurement of the polarization of initial-state photons in
polarized photon collisions at high energies typical of the
photon-photon collision mode of the ILC.  Thus, it is conceivable
that such technology could be extended to the needs of a detector
that would seek to resolve the final-state photon helicity. We
therefore advocate the
construction of  such photon helicity filtering detectors at the  ILC 
that can open up the possibility of further improving 
the  bound on {\rm Re}$\lambda_1$.

\section{Summary and Discussion}\label{summary}

In this paper, we have studied the process $e^+e^- \to \gamma Z$
at the ILC at $\sqrt{s}$= 500 GeV and 800 GeV with a realistic integrated luminosity.
We have pointed out the benefits of the resolution of final-state spin 
for studying the effects of couplings which were otherwise invisible with initial 
longitudinal as well as transverse states. Inspired by the work of final-state 
top-spin measurement we have shown that one can isolate 
${\rm Re}\lambda_1~(\gamma Z Z)$ in 
the above process  with initial transverse states by 
the measurement of the helicity of the $Z$ or the photon.

We have also made a numerical study of the limits on various couplings that
could be obtained at a future linear collider  assuming realistic transverse
polarizations of 80\% and 60\% respectively for $e^-$ and $e^+$, respectively.
We have also given the contour plot for allowed region in 
${\rm Re}\lambda_1$-${\rm Re}\lambda_2$ plane.  We see that with final-state
spin resolution, transverse
polarization can provide a sensitive test of the anomalous coupling,
{\rm Re}$\lambda_1$.  Overall it is seen that the resolution
of helicity of $\gamma$ gives better limits on the couplings
compared to the case when the helicity of $Z$ is resolved.
The above is due to the fact that only transversely polarized $Z$
is sensitive to {\rm Re}$\lambda_1$, but its contribution is
smaller by a factor of $(\bar{s}-1)$.  Moreover the contribution
of {\rm Re}$\lambda_1$ and {\rm Re}$\lambda_2$ to the asymmetries
$A^Z_+(\theta_0)$ and  $A'^Z_-(\theta_0)$  is inversely proportional to
$\sqrt{s}$. Therefore in the case $\sqrt{s}$ = 800 GeV the
resolving power of asymmetries using the helicity of the $Z$ is suppressed
compared to the case when the helicity of $\gamma$ is resolved. 
Since the anomalous couplings ${\rm Re}\lambda_1$ and ${\rm Re}\lambda_2$
cannot be completely disentangled through helicity measurements
we present both simultaneous and individual limits where the latter
is obtained by setting one of the couplings to zero at a time.
These limits are found to be stable, when fed back into the expressions
to quadratic order for the cross section. The asymmetries and the limits 
on the couplings are also calculated with the inclusion of quadratic terms
for the cross section. The new limits obtained are summarized  in
Tables~\ref{G_table1} and \ref{G_table2}.

It must, however, be mentioned that
one cannot directly isolate events with $Z$ helicities of $+1$ or $-1$. Hence
to measure the asymmetries we discuss, one would have to carry out a
subtraction of events in two kinematic regions of the decay products
corresponding to positive and negative polarizations of the $Z$. Doing
so would entail a loss of efficiency to a certain extent. We have not
taken this into account. It may be possible to consider the cases
when the helicities of $Z$ as well as $\gamma$ can be resolved.
Indeed, the technology with aligned crystals, see Ref.~\cite{Na59Collb}
could be adapted for the final-state photon helicity resolution.
It may be possible to carry out a study based on this,
but is beyond the scope of the present work, as the features that
we wish to study are already apparent when we sum over the helicity
of one of the other.  Furthermore, measuring both spins would lead 
to a loss in statistics thereby making this option less attractive.
Additional studies beyond the scope of the present work are related
to polarimetry.

We conclude by discussing some further approaches of
value to the process at hand.
It has  been proposed that
the hard radiative decay of the $Z$ can also be used to
place bounds on the magnitude of anomalous CP-violating
couplings~\cite{PRZ}.
Of special significance is the construction presented in Refs.~\cite{Ots1,Ots2}
of the formalism of a general spin-1 density matrix
for the $Z$-boson spin orientation introduced in the context
of CP--conserving anomalous couplings.  It would be of interest
to see if this can be extended to the case of CP--violating
anomalous couplings and so explore the possibility of
using this formalism to obtain bounds on such couplings
as we have done here.  
In Refs.~\cite{BASDRc1,BASDRc2}, the process has been studied
with BSM interactions given by most general contact interactions
with the helicities of the $\gamma$ and $Z$ summed over.  It 
may be of interest to apply the present considerations to 
this scenario as well.

\bigskip

\noindent {\bf{Acknowledgements:}} 
BA thanks the
Department of Science and Technology, Government of India,
and the Homi Bhabha Fellowships Council for support.
SDR thanks the Department of Science and Technology, Government of
India, for support under the J.C. Bose National Fellowship program,
grant no. SR/SB/JCB-42/2009.

\end{document}